\documentclass[pra,aps,twocolumn,showpacs]{revtex4-1}
\usepackage{amsmath,amssymb,graphicx}
\usepackage{epsfig}
\usepackage{textcomp}
\usepackage{color}
\usepackage{epstopdf}
\usepackage{hyperref}
\hypersetup{pdfpagemode=UseNone,colorlinks=true,linktoc=page,pdfborder={0 0 0},linkcolor=blue,citecolor=blue}

\begin{document}

\title{Capture into resonance and phase space dynamics in optical centrifuge}
\author{Tsafrir Armon and Lazar Friedland}
\affiliation{Racah Institute of Physics, Hebrew University
of Jerusalem, Jerusalem 91904, Israel}
\email{lazar@mail.huji.ac.il}

\begin{abstract}
The process of capture of a molecular enesemble into rotational resonance in
the optical centrifuge is investigated. The adiabaticity and phase
space incompressibility are used to find the resonant capture probability in
terms of two dimensionless parameters $P_{1,2}$ characterising the driving
strength and the nonlinearity, and related to three characteristic time
scales in the problem. The analysis is based on the transformation to
action-angle variables and the single resonance approximation, yielding
reduction of the\ three-dimensional rotation problem to one degree of
freedom. The analytic results for capture probability are in a good
agreement with simulations. The existing experiments satisfy the validity
conditions of the theory.
\end{abstract}

\maketitle


\section{Introduction}

\label{introduction} The field of optical control and manipulation of
molecular rotation has seen major advances over the years, and today various
techniques allow to control the rotation alignment \cite%
{alignment1,alignment2}, orientation \cite{orientation1,orientation2} and
directionality \cite{rotation1,rotation2,rotation3} of molecular ensembles.
One of the most innovative tools in this field is the \textit{optical
centrifuge} (OC), originally proposed and implemented by Corkum and
collaborators \cite{Corkum1,Corkum2}, who introduced the possibility of
controlled excitation of the molecular rotational degree of freedom by
chirped laser pulses. The controlled nature of this process is twofold: the
molecules reach very high rotational states (\textit{super rotors}), but
they also remain closely centered around a specific target energy/frequency.
The controlled rotation could be used to selectively dissociate molecules
\cite{Corkum2} or a specific molecular bond \cite{HCNbreaking} and has been
shown to change molecular characteristics, such as the molecule's stability
against collisions \cite{Forrey} and its scattering from surfaces \cite%
{averbukh1}. Furthermore, a gas of super rotors may exhibit new optical
properties \cite{averbukh2} and formation of vortices \cite{averbukh3}.

Over the last few years, several state of the art experiments have been
performed \cite{mullin1,milner1,milner2} utilizing different molecules, and
exploring the dynamics during and after the OC laser pulse, including the
excitation process \cite{milner1}, the gyroscopic stage in which the
molecules remain oriented \cite{milner3} and the equilibration and
thermalization that follows the pulse and produces an audible sound wave
\cite{milner4}. However, while the experimental setups improved
considerably, the process of capture of molecules into the chirped resonant
rotation is still poorly understood. This process was only studied
numerically \cite{ivanov1} or under the constraint that the molecules rotate
in a plane perpendicular to the laser propagation axis \cite%
{Corkum1,ivanov2,girard}. The former asumption makes it impossible to study
the response of a randomly oriented molecular ensemble to the OC pulse. As a
result, the efficiency of the OC, i.e. the fraction of molecules captured by
the chirped laser drive, was not analyzed sufficiently.

In this work, we will show that under the rigid-rotor approximation the OC
is an example within a broad family of driven non-linear systems exhibiting
a sustained phase-locking or \textit{autoresonance} (AR) with a chirped
drive. This phenomenon has been observed and studied in many applications,
including atomic systems \cite{AR1,AR2}, plasmas \cite{AR3,AR4}, fluids \cite%
{AR5}, and semiconductor quantum wells \cite{AR6}. By using methods in the
theory of AR and analyzing the associated phase space dynamics we will for
the first time calculate the efficiency of the OC process. The quantum
counterpart of the AR is the quantum energy ladder climbing \cite%
{AR6a,AR7,AR8}, but we will show that the classical AR analysis is relevant
to many current experimental setups.

The scope of the paper will be as follows. In Sec. \ref{model}, we will
discuss the driven-chirped molecular rotation in three dimensions, transform
to action-angle variables, and use the single resonance approximation to
reduce the problem to one degree of freedom. Section \ref{TE} will focus on
calculating the efficiency of the resonant capture process in the system via
analysing its dynamics in a continuous phase space instead of a single
particle approach. In section \ref{ResDis}, we will compare the theory with
numerical simulations and discuss the validity of our approximations and the
applicability to current experimental setups. Our conclusions will be
summarized in Sec. \ref{summary}.

\section{The model}

\label{model}

\subsection{Parameterization}

The fundamental idea of the OC, is that an anisotropic molecule will "chase"
(and, thus, be rotationally excited) a rotating linearly polarized wave,
whose polarization rotation accelerates over time. In practice, such driving
wave is created by combining two counter rotating and antichirped circularly
polarized laser beams \cite{Corkum1}. For a wave propagating along the $Z$
axis, with polarization angle $\phi _{d}\left( t\right) $ in the $XY$ plane,
after averaging over the optical frequency of the laser beams, the
interaction potential energy of a molecule in spherical coordinates is given
by $U=-\varepsilon \sin ^{2}\theta \cos ^{2}\left( \varphi -\phi _{d}\right)
$ \cite{Corkum1}, where $\varepsilon =\left( \alpha _{\parallel }-\alpha
_{\perp }\right) E_{0}^{2}/4$, $\alpha _{\parallel }$,$\alpha _{\perp }$ are
the polarizability components of the molecule and $E_{0}$ is the electric
field amplitude of the combined beam. For simplicity, we will use a \textit{%
linearly chirped} driving frequency $\omega _{d}=d\phi _{d}/dt=\beta t/2$,
where $\beta >0$ is the chirp rate, but any sufficiently slow chirp will
lead to similar results. The initial rotation frequency is set by taking an
appropriate intial time.

Our driven system can be characterized by three different time scales, i.e.
the drive sweeping time $t_{s}=1/\sqrt{\beta }$, the characteristic thermal
rotation time $t_{th}=1/\omega _{th}=\sqrt{I/k_{B}T}$, and the driving time
scale $t_{d}=L_{th}/\varepsilon =\sqrt{Ik_{B}T}/\varepsilon $, where $T$ is
the temperature, $I$ the molecule's moment of inertia, and $L_{th}=I\omega
_{th}$ is the characteristic thermal angular momentum. These three time
scales define two dimensionless parameters,
\begin{equation}
P_{1}=\frac{t_{s}}{t_{d}}=\frac{\varepsilon }{\sqrt{Ik_{B}T\beta }},
\label{P1}
\end{equation}
which measures the drive's strength, and
\begin{equation}
P_{2}=\frac{t_{s}}{t_{th}}=\sqrt{\frac{k_{B}T}{I\beta }},  \label{P2}
\end{equation}
characterizing the nonlinearity of the problem. These parameters enter
naturally in the dimensionless Hamiltonian of our driven system in spherical
coordinates
\begin{equation}
H=\frac{P_{2}}{2}\left( p_{\theta }^{2}+\frac{p_{\varphi }^{2}}{\sin ^{2}{%
\theta }}\right) -P_{1}\sin ^{2}\theta \cos ^{2}\left( \varphi -\phi
_{d}\right) ,  \label{H-sphe}
\end{equation}%
where we normalize the canonical momenta and later the total angular
momentum $L$ with respect to $L_{th}$, and use the dimensionless time $\tau =%
\sqrt{\beta }t$.
\begin{figure}[t]
\includegraphics[width=3.375in,clip=]{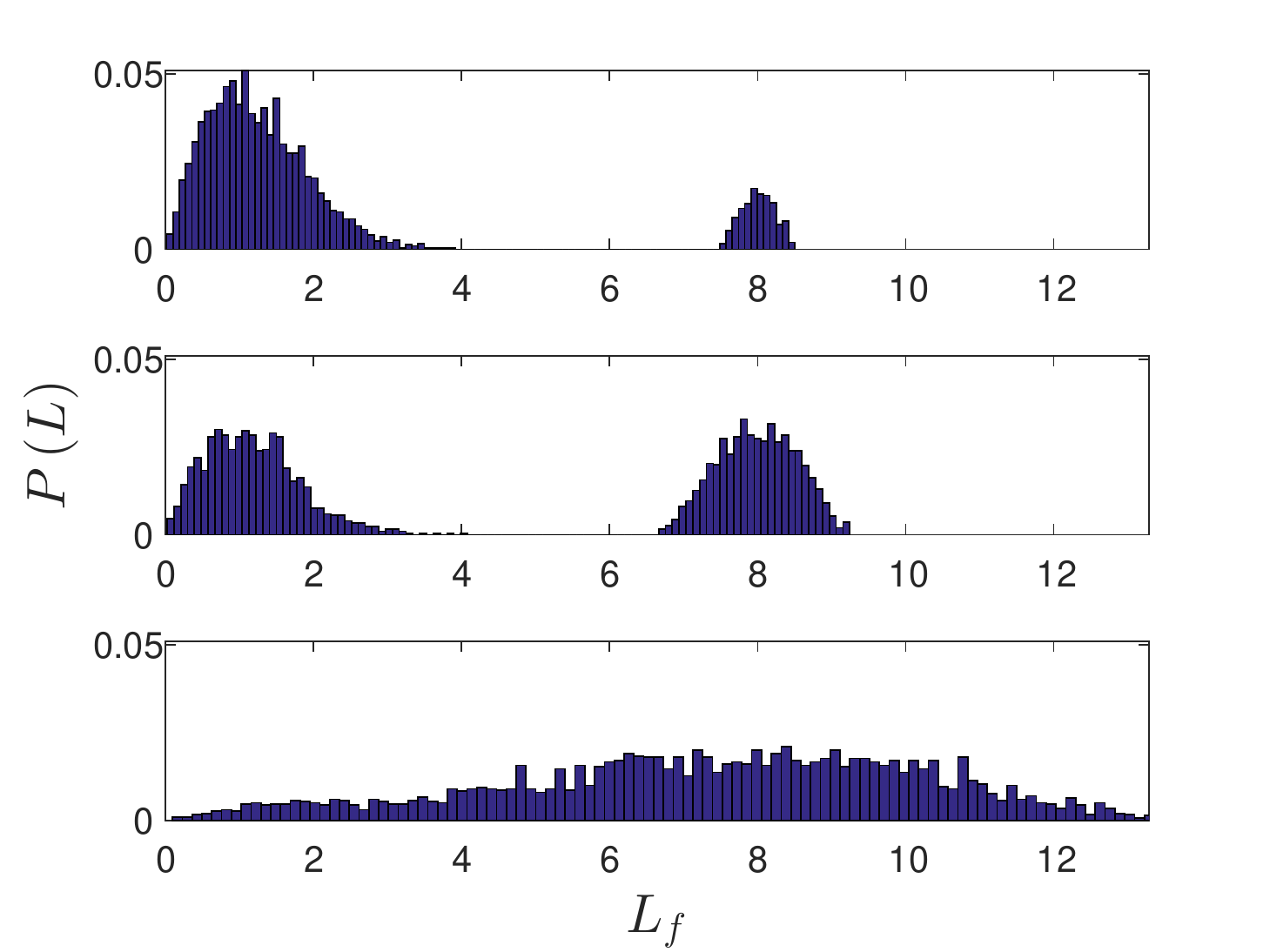}
\caption{Monte Carlo simulation of the distribution of angular momenta $%
L_{f} $ for initially thermal ensemble (3000 molecules) after the OC pulse
with initial and final normalized driving frequencies $\protect\omega _{0}=1$
and $\protect\omega _{f}=8$. The oarameters are $P_{2}=2.51$ and $%
P_{1}=0.63,2.51,39.8$ in pannels (a)-(c), respectively}
\label{fig-a}
\end{figure}
The evolution equations based on this Hamiltonian comprised one of the two
sets used for Monte Carlo simulations in this work. Figure \ref{fig-a}
shows the distributions (histograms) of the normalized angular momenta at
the end of the chirped OC drive after starting from initially thermal
molecular ensemble. The resonant normalized angular momentum in the OC
equals the instantaneous driving frequency normalized with respect to $%
\omega _{th}$ (see below). The initial and final normalized driving
frequencies in Fig. \ref{fig-a} were $1$ and $8$, respectively, and we used
parameters $P_{2}=2.51$ and $P_{1}=0.63$ (Fig. 1a), $2.51$ (Fig. 1b), $39.8$
(Fig. 1c). When parameter $P_{1}$ is increased (for constant $P_{2}$ this
corresponds to increasing the laser intensity), more molecules experience
significant acceleration. Nevertheless, if one seeks a narrow distribution
around a specific target frequency, the acceleration in panels (b) and (c)
Fig. \ref{fig-a} does not provide the desired level of control, showing a
broad distribution around the target. In contrast, panel (a), is a
representative example for the degree of control and accuracy one can
achieve with the OC, provided the parameters are chosen appropriatly. In
this work we calculate the excitation efficiency and the width of the final
distribution of the angular momentum in the $P_{1,2}$ parameter space.

\subsection{Transformation to action-angle variables and single-resonance
approximation}

\label{AAtransform} Like in many other physical systems, it is convenient to
transform our driven problem to the action-angle variables of the
unperturbed problem since the latter is integrable. This canonical
transformation $\theta ,\varphi ,p_{\theta },p_{\varphi }\rightarrow \Theta
_{L},\Theta _{L_{z}},L,L_{z}$ (see Appendix A for details) leads to non
trivial angle variables (related to Euler angles), while the actions $L$ and
$L_{z}$ are the normalized total angular momentum and its projection on the $%
Z$ axis. The transformed Hamiltonian assumes the form:
\begin{equation}
H\left( \Theta _{L},\Theta _{L_{z}},L,L_{z}\right) =P_{2}\frac{L^{2}}{2}%
+P_{1}U\left( \Theta _{L},\Theta _{L_{z}},L_{z}/L,\phi _{d}\right) ,
\label{H-AA}
\end{equation}%
where $U$ is a periodic function of $\Theta _{L},\Theta _{L_{z}}$ of period $%
\pi $, and its exact form is presented in the appendix.

\begin{figure}[t]
\includegraphics[width=3.375in,clip=]{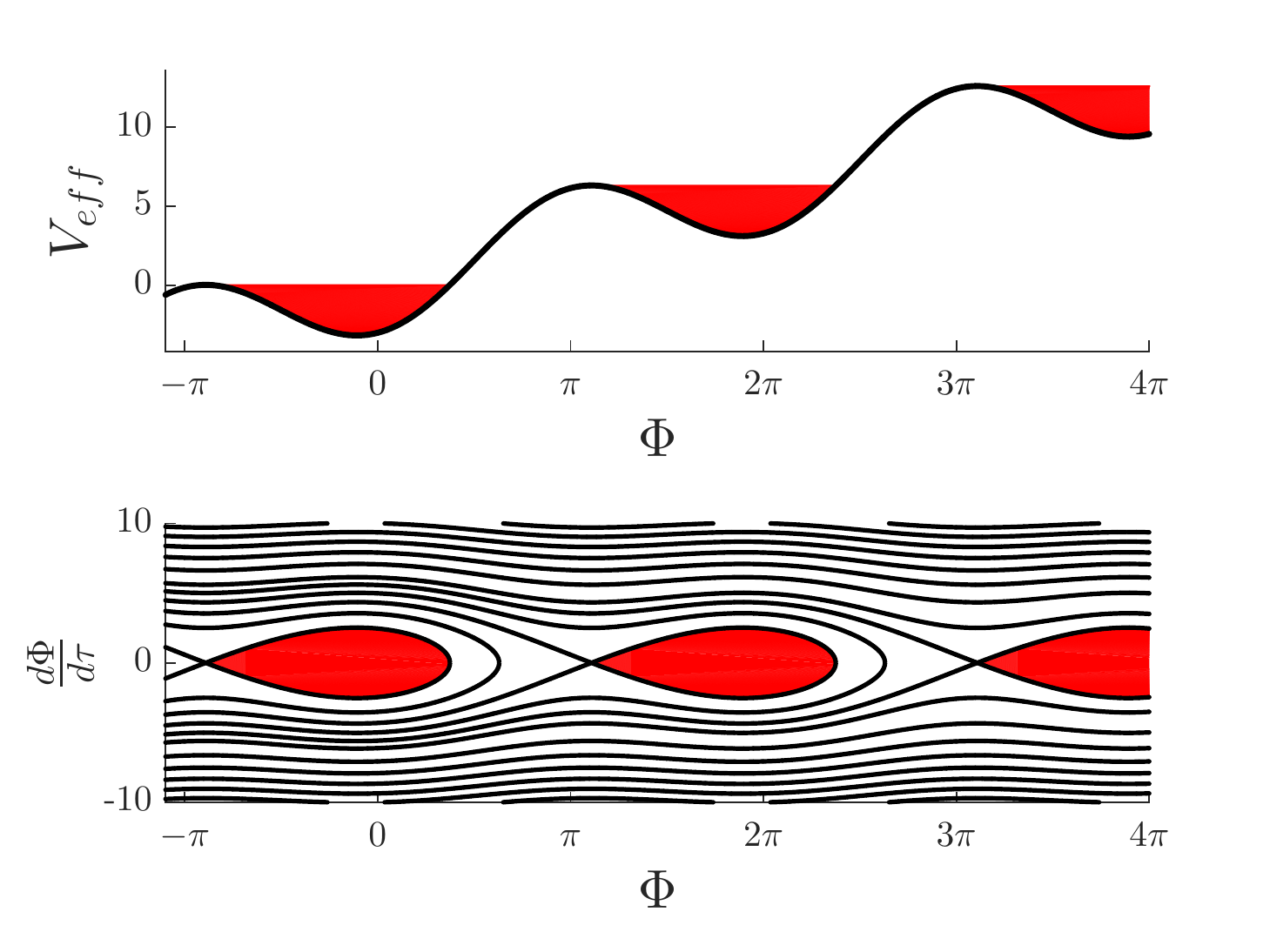}
\caption{The effective potential (Eq. \protect\ref{Veff}) (top panel) and
the phase-space portrait of the associated dynamics (bottom panel). The
boundary of the red filled area in the bottom panel is the separatrix. The
value of $P_{1}P_{2}V$ was $0.75$ and the equal energy lines in the bottom
panel are separated by energy steps of $\protect\pi $.}
\label{fig-j}
\end{figure}

The perturbing part in (\ref{H-AA}) contains several oscillating terms,
however, the main resonance in our case is defined by requiring stationarity
$\Phi \approx const$ of the phase-mismatch $\Phi =2(\Theta _{L}+\Theta
_{L_{z}}-\phi _{d})$. Assuming a weak drive, i.e. $P_{1}/P_{2}\ll 1$ (this
approximation will be discussed in Sec. IV) in the vicinity of the
resonance, we can use the single resonance approximation \cite{chirikov},
i.e. discard all the rapidly oscillating terms in the Hamiltonian. The
resulting approximate, single resonance Hamiltonian is (see Appendix A):
\begin{equation}
H_{r}=P_{2}\frac{L^{2}}{2}+P_{1}V\cos {\Phi }+P_{1}F,  \label{HAASR}
\end{equation}%
where
\begin{eqnarray}
V &=&\frac{1}{8}\left( 1+\frac{L_{z}}{L}\right) ^{2}, \\
F &=&\frac{1}{4}\left( 1-\frac{{L_{z}}^{2}}{L^{2}}\right) .
\end{eqnarray}%
The corresponding evolution equations are
\begin{eqnarray}
\frac{d\Theta _{L}}{d\tau } &=&P_{2}L-P_{1}\frac{L_{z}}{L^{2}}\left(
V^{\prime }\cos {\Phi }+F^{\prime }\right) ,  \label{SR3} \\
\frac{d\Theta _{L_{z}}}{d\tau } &=&P_{1}\frac{1}{L}\left( V^{\prime }\cos {%
\Phi }+F^{\prime }\right) ,  \label{a} \\
\frac{dL}{d\tau } &=&2P_{1}V\sin {\Phi },  \label{b} \\
\frac{dL_{z}}{d\tau } &=&2P_{1}V\sin {\Phi }.  \label{c}
\end{eqnarray}%
Here, the prime denotes differentiation with respect to $L_{z}/L$. Equations
(\ref{b}), (\ref{c}) yield the integral of motion $C=L-L_{z}$ ($0\leq C\leq
2L$), which allows reduction to a single degree of freedom:
\begin{eqnarray}
\frac{dL}{d\tau } &=&2P_{1}V\sin {\Phi },  \label{SR4} \\
\frac{d\Phi }{d\tau } &=&2P_{2}L+2P_{1}\frac{C}{L^{2}}\left( V^{\prime }\cos
{\Phi }+F^{\prime }\right) -\tau .  \label{SR4a}
\end{eqnarray}%
Equation (\ref{a}) still needs to be solved to obtain the precession of the
angular momentum around the $Z$-axis, but for calculating $L,$ the one
degree of freedom set above is sufficient. This is our second (approximate)
set used in the simulations below, which, due to the adiabaticity and
reduced number of degrees of freedom, is considerably faster numerically
than the full set of evolution equations in terms of the original spherical
coordinates. We will assume, and verify a posteriori that if $\Delta L\ll 1$
is the range of $L$ values in a persistent resonance with the drive in our
problem, then $P_{2}\Delta L\gg P_{1}$. Under this assumption, the second
term in Eq. (\ref{SR4a}) can be neglected, and the phase locking (resonance)
condition $d\Phi /d\tau \approx 0$ yields $2P_{2}L-\tau \approx 0$. Let $%
L_{r}(\tau )=\tau /2P_{2}=\omega _{d}(\tau )/\omega _{th\text{ }}$be the
value of $L$ satisfying the resonance condition exactly and define the
deviation $\delta L=L-L_{r}$ from the exact resonance. The evolution
equations then yield
\begin{eqnarray}
\frac{d\delta L}{d\tau } &=&2P_{1}V\sin {\Phi }-\frac{1}{2P_{2}},
\label{SR6} \\
\frac{d\Phi }{d\tau } &=&2P_{2}\delta L.  \label{SR6a}
\end{eqnarray}%
By taking the derivative of Eq. (\ref{SR6a}) with respect to time and
inserting Eq. (\ref{SR6}), we get
\begin{equation}
\frac{d^{2}\Phi }{d\tau ^{2}}=-4P_{1}P_{2}V\sin {\Phi }-1,  \label{SR7}
\end{equation}%
where we shifted $\Phi $ by $\pi $ and, to lowest order in $\delta L$, $%
V\approx \frac{1}{8}\left( 2-\frac{C}{L_{r}}\right) ^{2}$ is evaluated at $%
L_{r}$. Equation (\ref{SR7}) describes a pseudo-pendulum under the action of
a constant torque. The Hamiltonian in this problem, with $d\Phi /d\tau $
acting as the momentum, is:
\begin{equation}
H=\frac{1}{2}\left( \frac{d\Phi }{d\tau }\right) ^{2}+V_{eff}\left( \Phi
\right) ,  \label{H}
\end{equation}%
where
\begin{equation}
V_{eff}\left( \Phi \right) =-4P_{1}P_{2}V\cos {\Phi }+\Phi .  \label{Veff}
\end{equation}%
This tilted cosine effective potential and the associated phase space
portrait of dynamics of the pseudo-pendulum are shown in Fig. \ref{fig-j}
for $P_{1}P_{2}V=0.75$. The phase space (bottom panel in the figure) is
comprised of open and closed trajectories, provided $P_{1}P_{2}V>1/4$. The
open trajectories exhibit a continuous growth of the phase-mismatch, i.e.
are not phase locked with the drive, while for the closed trajectories the
phase-mismatch is bounded. The closed trajectories are surounded by the
separatrix having area shown in red in the bottom panel of the figure. As $%
V(L_{r})$ in our problem is slowly varying (increasing) in time, both the
closed and open trajectories evolve adiabatically in time, unless near the
separatrix. This means that deeply trapped trjectories remain trapped, i.e
the rotation frequency follows the drive, $L(\tau )\approx \omega _{d}(\tau
)/\omega _{th\text{ }}$, constituting the AR in the system. The main problem
remains the fate of the trajectories near the separatrix. These
trajectories, in principle, can change their trapping status as the result
of nonadiabatic dynamics and, thus, \ affect the OC efficiency. It should be
mentioned that many other AR systems \cite{AR3,AR4,AR5,AR6} are described by
the resonant Hamiltonian similar to (\ref{H}). The process of capture into
resonance in all these problems depends critically on the specific form of
function $V$. In many such problems $V\sim \sqrt{I}$, where $I$ is the
relevant action variable in the problem. In all such cases, the capture into
resonance from equilibrium and transition to AR is guaranteed provided the
driving amplitude exceeds a sharp threshold \cite{AR3}. Because of a
different dependence of $V$ on $L$ no such threshold is characteristic of
the driven molecule case. The study of this different capture mechanism
comprises the main goal of the present investigation.


\section{\protect\bigskip Trapping Efficiency}

\label{TE}

\subsection{The complexity of resonant trapping problem}

\begin{figure}[t]
\includegraphics[width=3.375in,clip=]{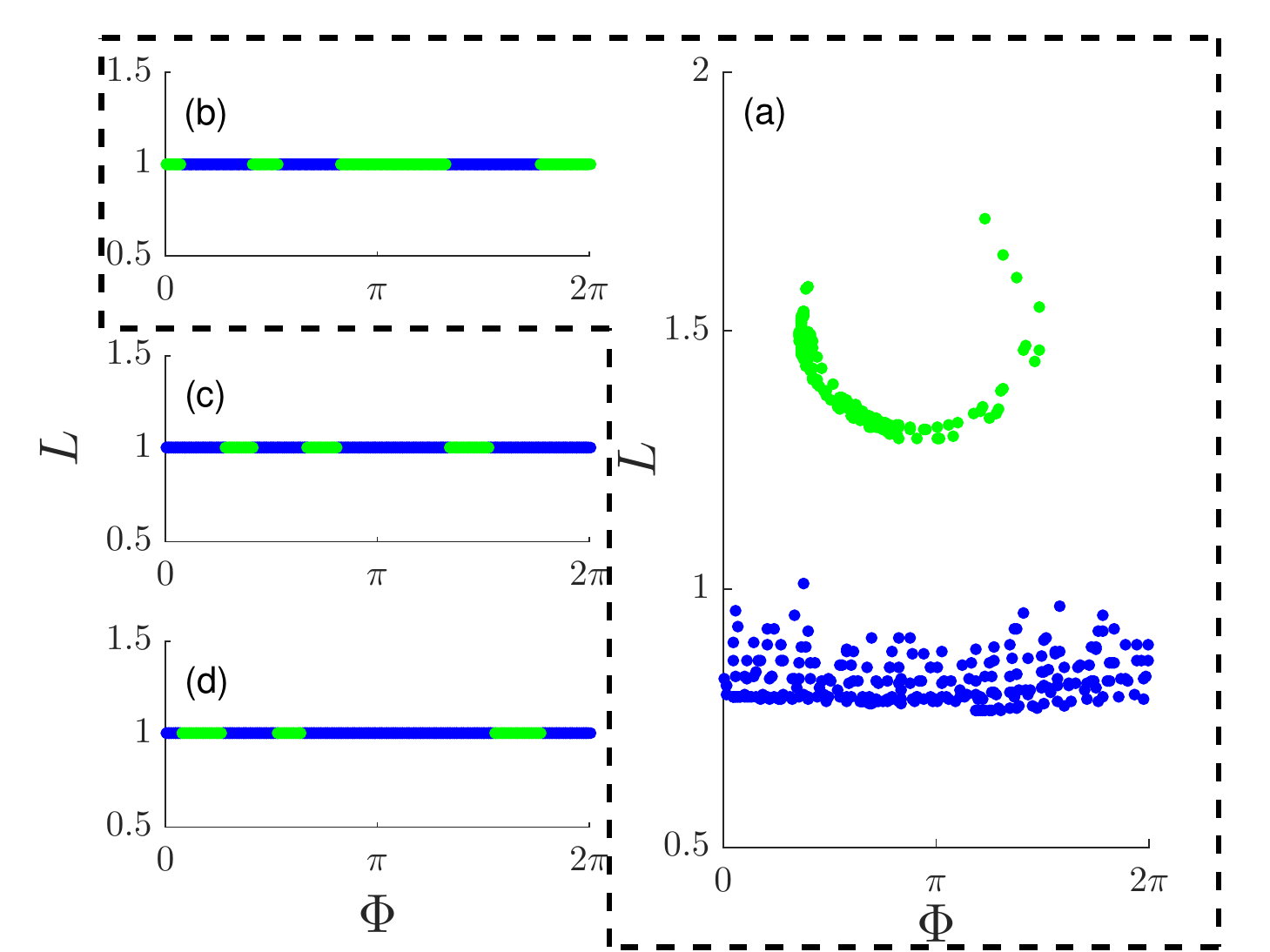}
\caption{(Color online) Numerical simulations (single resonance
approximation) of passage through resonance with 500 molecules with initial $%
L_{0}=1$ and $C=1$. All panels show the phase-space ($\Phi $,$L$), with
green and blue circles representing resonantly trapped and untrapped
molecules, respectively. The left panels show the initial distributions and
differ by a small shift of the initial driving frequency: $\protect\omega %
_{0}=0.5$ (b), $0.49$ (c) and $0.51$ (d). The right panel (a) shows the
final distribution of the initial condition (b) at $\protect\omega _{f}=1.5$%
. The parameters are $P_{1}=1$, $P_{2}=10$ and $\Phi $ is shifted so that $%
\Phi =0$ is at the saddle point (see subsection \protect\ref{PSD}).}
\label{fig-b}
\end{figure}

We have seen in simulations in Sec. \ref{model} that for a range of
parameters, the OC yields controlled rotational excitation of molecular
ensembles. Here we study the efficiency of such excitation process, i.e.
evaluate the fraction of molecules from some initial distribution, which are
captured into and remain in resonance. Intuitively, one can assume that if
the value of $V$ changes adiabatically, molecules will be either trapped or
not according to their initial location in phase space - inside or outside
the separatrix. While the changes of $V$ are generally adiabatic (as will be
seen later), this intuition proves to be wrong. Indeed, the molecules which
are inside the separatrix initially remain in resonance at later times, but
additional molecules can cross the separatrix and enter the trapped region
even if they were outside initially. An illustration of this process is
presented in Fig. \ref{fig-b}, where panel (a) shows the final phase-space
distribution of a molecular ensemble having the same $L=1$ and $C=1$
initially and uniformly distributed values of $\Phi $ (see panel b). The
normalized driving frequency was varied from $\omega _{0}=0.5$ to $\omega
_{f}=1.5$, and one can see that despite a much lower initial driving
frequency compared to the rotation frequency of the molecules, a
considerable amount of molecules end up captured into resonance and
rotationally accelerated (green). The location of the newly trapped
molecules in the initial ensemble is shown in green in panel (b). We find
that this location and the fraction of trapped molecules strongly depends on
the initial value of the driving frequency. This complexity is illustrated
in panels (c) and (d), showing in green the location of the molecules
trapped in resonance with the drive for the same initial conditions, but
with the initial driving frequency changed by $\pm 0.01$. The fraction of
the trapped molecules in cases (c) and (d) was $22\%$ and $26\%$,
respectively compared to $46\%$ in the case (a-b).

One approach to deal with the nonadiabatic passage through separatrix
problem is to study an ensemble of initial conditions, checking whether the
associated trajectories cross the separatrix. Previous works used such
approach with simpler systems, but the probabilistic nature of this
nonadiabatic phenomenon led to rather complex results \cite%
{dioctorn,neishtadt}. Here, we will develop an alternative approach which
examines the continuous phase space dynamics of the initial ensemble,
instead of working with a collection of individual trajectories. This
approach will yield the resonant capture probability, without ever
specifying which initial conditions yield trajectories crossing the
separatrix.

\subsection{Phase-space dynamics}

\label{PSD} 
\begin{figure}[t]
\includegraphics[width=3.375in,clip=]{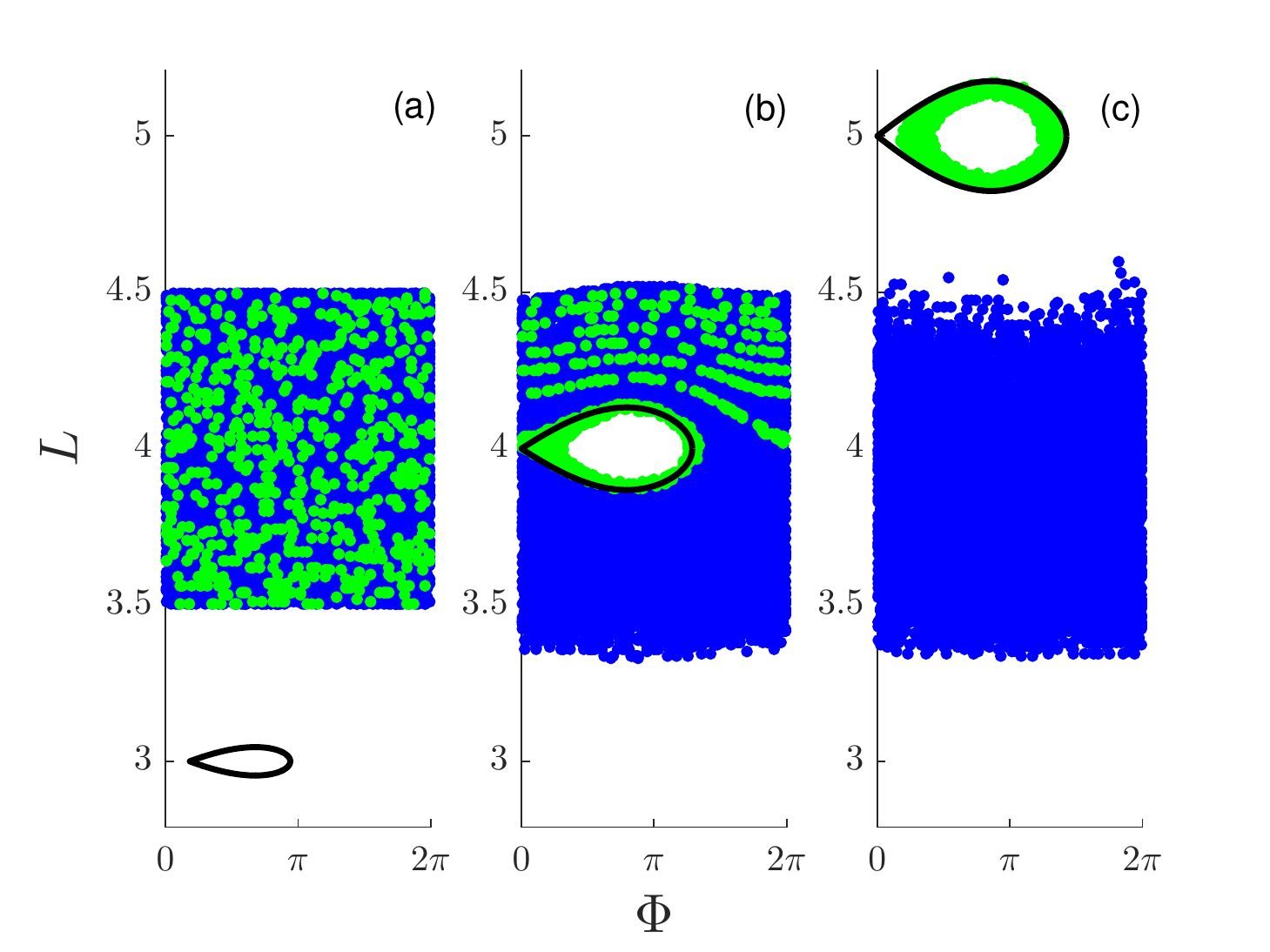}
\caption{(Color online) Numerical simulations (single resonance
approximation) for $10^4$ molecules distributed uniformly between $L=3.5$
and $4.5$ initially with $C=4$. The panels show the distribution of the
ensemble at three consecutive times (in terms of the normalized driving
frequency): $\protect\omega _{d}=3$ (a),$4$ (b) and $5$ (c). Green and blue
circles show resonantly trapped and untrapped molecules, respectively, and
the black lines are the associated separatrixes. The parameters are $%
P_{1}=0.63$ and $P_{2}=10$. $\Phi$ is shifted so that $\Phi =0$ is at the
saddle point of panel (c). A video of the simulation is provided in the
online supplementary material \cite{movie}.}
\label{fig-c}
\end{figure}

We base our analysis on Eq. (\ref{SR7}), where $V=\frac{1}{8}%
\left( 2-\frac{C}{L}\right) ^{2}$ is evaluated at $L_{r}$ and, therefore,
both $V$ and the associated separatrix area are monotonically increasing
functions of time. For molecules close to the separatrix, trapped or
untrapped, this approximation is satisfied because $\Delta L\ll 1$, where
now we associate $\Delta L$ with the width of the separatrix in $L$. For
untrapped molecules far from the separatrix we can still evaluate $V$ at $%
L_{r}$, because the phase mismatch $\Phi $ for such molecules varies
rapidely and the effect of the driving term in the quasipotential averages
out. Next, instead of passage through resonance with an ensemble of
molecules having the same value of $L$ as illustrated in Fig. \ref{fig-b},
we consider an ensemble of molecules with initially uniform density in phase
space between $L_{1}=3.5$ and $L_{2}=4.5$ with all molecules having the same
$C=4$. We show a numerical simulation in such a system as the driving
frequency (and therefore $L_{r}$) successively passes the resonance with all
the molecules in the ensemble in Fig. \ref{fig-c} (a video of this
simulation can be found in the online supplementary material \cite{movie}). As the
driving frequency sweeps through the ensemble [time progresses from (a) to
(c)], the area of the associated separatrix (in black) increases and the
added area is filled with the same density of molecules as in the original
distribution. However, the area of the separatrix which was empty when the
separatrix first entered the distribution, remains empty forming a phase
space hole passing through the distribution (similar phase space holes were
studied in plasma physics applications \cite{pavel}). Note that the
separatrix crossing occurs near the saddle point (where the adiabacity
condition is not met) and the molecules "line up" to enter the separatrix,
as seen in panel (b) in the figure. Following the crossing, the area filled
by the newly trapped molecules is very regular, and the only irregular
regions of phase space after passage through resonance are those near the
boundaries $L_{1,2\text{ }}$ of the original distribution. Furthermore, one
can observe that the whole distribution is shifted to lower values of $L$
after the drive completed its passage through the ensemble.

The resonant phase space dynamics shown in Fig. \ref{fig-c} can be explained
on the bases of (a) the adibaticity in the problem \cite{landau} and (b) the
incompressibility of the phase space \cite{Goldstein-inco}. The adiabaticity
guarantees the conservation of the area of the empty hole inside the growing
separatrix, while the incompressibility of the phase space ensures that the
distribution of the newly trapped molecules inside the separatrix would be
the same uniform (original) distribution as long as $L_{r}$ is well within
the range $L_{1},L_{2}$. Therefore, as time progresses and the resonant
separatrix passes an infinitesimal distance $\delta L_{r}$ inside the
distribution, the density $\delta N$ of newly trapped molecules is
\begin{equation}
\delta N=P\delta S=P\frac{\partial S}{\partial L_{r}}\delta L_{r},
\label{dN}
\end{equation}%
where $P$ is the initial (uniform) density of the molecules in phase space
and $\delta S$ is the change of the area of separatrix during the
corresponding infinitesimal time interval. Thus, the number of the newly
trapped molecules after passage through the whole distribution is $\Delta
N=P\Delta S$, $\Delta S$ being the full added area of the separatrix after
the passage. These simple arguments also allow us to calculate the
probability of capture into resonance for a general initial distribution of $%
L$ and $C$, (i.e. $L_{z}$), which will be discussed next.

\subsection{Capture Probability}

The generalization to the case of an arbitrary initial phase space density
distribution $P(L,C)$ independent of $\Phi $ can proceed by viewing this
distribution as a collection of uniform infinitesimally thin layers, each
having some value of $C$. As the most prevailing case, we focus on initially
thermal distribution of molecules, where the distribution of $L$ is
\begin{equation}
P_{th}\left( L\right) =L\exp {\left( -\frac{L^{2}}{2}\right) },  \label{TD-a}
\end{equation}%
and, therefore,
\begin{equation}
P(L,C)=\left\{
\begin{array}{ccc}
0 & , & L<C/2 \\
\frac{NP_{th}\left( L\right) }{4\pi L} & , & L>C/2%
\end{array}%
\right. ,  \label{Dis}
\end{equation}%
\begin{figure}[t]
\includegraphics[width=3.375in,clip=]{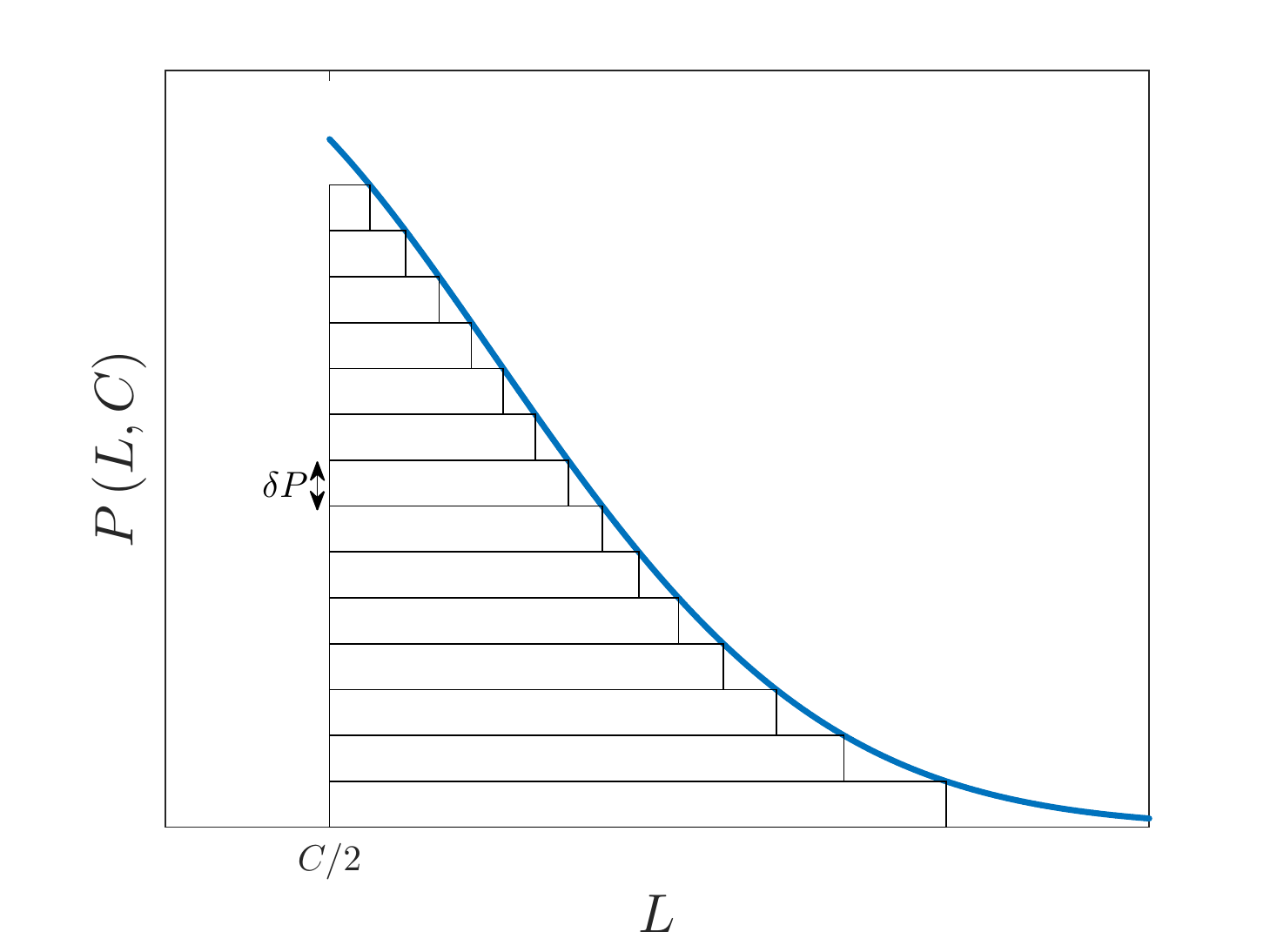}
\caption{Phase space density distribution $P(L,C)$ viewed as a collection of
uniform layers of height $\protect\delta P$ each.}
\label{fig-d}
\end{figure}
where $N$ is the density of the molecules. For a given $C$, we view this
distribution as a collection of uniform layers of thickness $\delta P$ as
illustrated in Fig. \ref{fig-d}. The resonant drive passes all these layers,
so at any given time, we have a collection of identical separatrices around
the resonant $L_{r}$. Since the layers have a uniform density, and $\Delta
L\ll 1$, the passage of the separatrix through the layers can be treated as
discussed above. As the separatrix advances an infinitesimal distance $%
\delta L_{r}$, the total density (after summation over all the layers and
integration over $\Phi $) of newly trapped molecules for given $C$ will be
[see Eq. (\ref{dN})]

\begin{equation}
\delta N\left( L_{r},C\right) =P(L_{r},C)\frac{\partial S}{\partial L_{r}}%
\delta L_{r}.  \label{delP}
\end{equation}%
Next, we integrate (\ref{delP}) over $C$ and change the integration from $C$
to $R=L_{z}/L=1-C/L_{r}$, which is uniformly distributed between $-1,1$ to
get
\begin{equation}
\delta N(L_{r})=\delta L_{r}\frac{NP_{th}\left( L\right) }{4\pi L}%
\int_{-1}^{1}\frac{dS}{dR}\left( 1-R\right) dR.  \label{TD-1}
\end{equation}%
Finally, we collect the newly trapped molecules as the resonant $L_{r}$
passes from some initial $L_{r0}$ to a final value $L_{rf}$ (the normalized
driving frequency varies from $\omega _{0}$ to $\omega _{f}$) to get the
density of all newly resonantly trapped molecules
\begin{equation}
\Delta N=\int_{L_{r0}}^{L_{rf}}\frac{NP_{th}\left( L\right) }{4\pi L}%
dL_{r}\int_{-1}^{1}\frac{dS}{dR}\left( 1-R\right) dR.  \label{delN}
\end{equation}%
After integrating in $R$ (by parts) and in $L_{r}$, the last expression
becomes%
\begin{equation}
\Delta N=\sqrt{\frac{\pi }{2}}\frac{NQ}{4\pi }\left[ erf\left( \chi
_{f}\right) -erf\left( \chi _{0}\right) \right] ,  \label{delna}
\end{equation}%
where $\chi =L_{r}/\sqrt{2}=\omega _{d}/\left( \sqrt{2}\omega _{th}\right) $
and $Q=\int_{-1}^{1}SdR$ is the total "volume" of the separatrix in the
3-dimensional extended phase-space which includes the $R$ dimension. To get
the total density of trapped molecules, we must add the density of the
initially trapped molecules, which, for $\Delta L\ll 1$ is

\begin{equation}
\Delta N_{0}=\frac{QN}{4\pi}P_{th}\left( \omega _{0}\right) \text{.}
\end{equation}%
Then the total capture probability in the problem is

\begin{equation}
P_{cap}=\frac{Q}{4\pi }\left\{ \sqrt{\frac{\pi }{2}}\left[ erf\left( \chi
_{f}\right) -erf\left( \chi _{0}\right) \right] +P_{th}\left( \omega
_{0}\right) \right\} .  \label{Pcap}
\end{equation}%
Finally, $Q$ in the last equation can be found numerically via
\begin{equation}
Q=\frac{\sqrt{2}}{P_{2}}\int_{-1}^{1}dR\int_{\Delta \Phi }\sqrt{D\left(
1-\cos {\Phi }\right) +\sin {\Phi }-\Phi }d\Phi ,  \label{Q}
\end{equation}%
where $D=\sqrt{[4P_{1}P_{2}V\left( R\right) ]^{2}-1}$, $\Delta \Phi $ is the
width of the separatrix in $\Phi $, and we shifted $\Phi $ in (\ref{Q}) so
that $\Phi =0$ is at the saddle point. Note that $Q$ depends on $P_{2}$ and
the product $P_{1}P_{2}$ and, therefore, for a given $\omega _{0},\omega
_{f} $, the capture probability scales with temperature as $T^{-1/2}$ via $%
P_{2}$. Furthermore, asymptotically for large $P_{1}P_{2}$, $Q$ $\sim \sqrt{%
P_{1}/P_{2}}$, which is independent of the chirp rate $\beta $.


\section{Results and Discussion}

\label{ResDis}

\begin{figure}[t]
\includegraphics[width=3.375in,clip=]{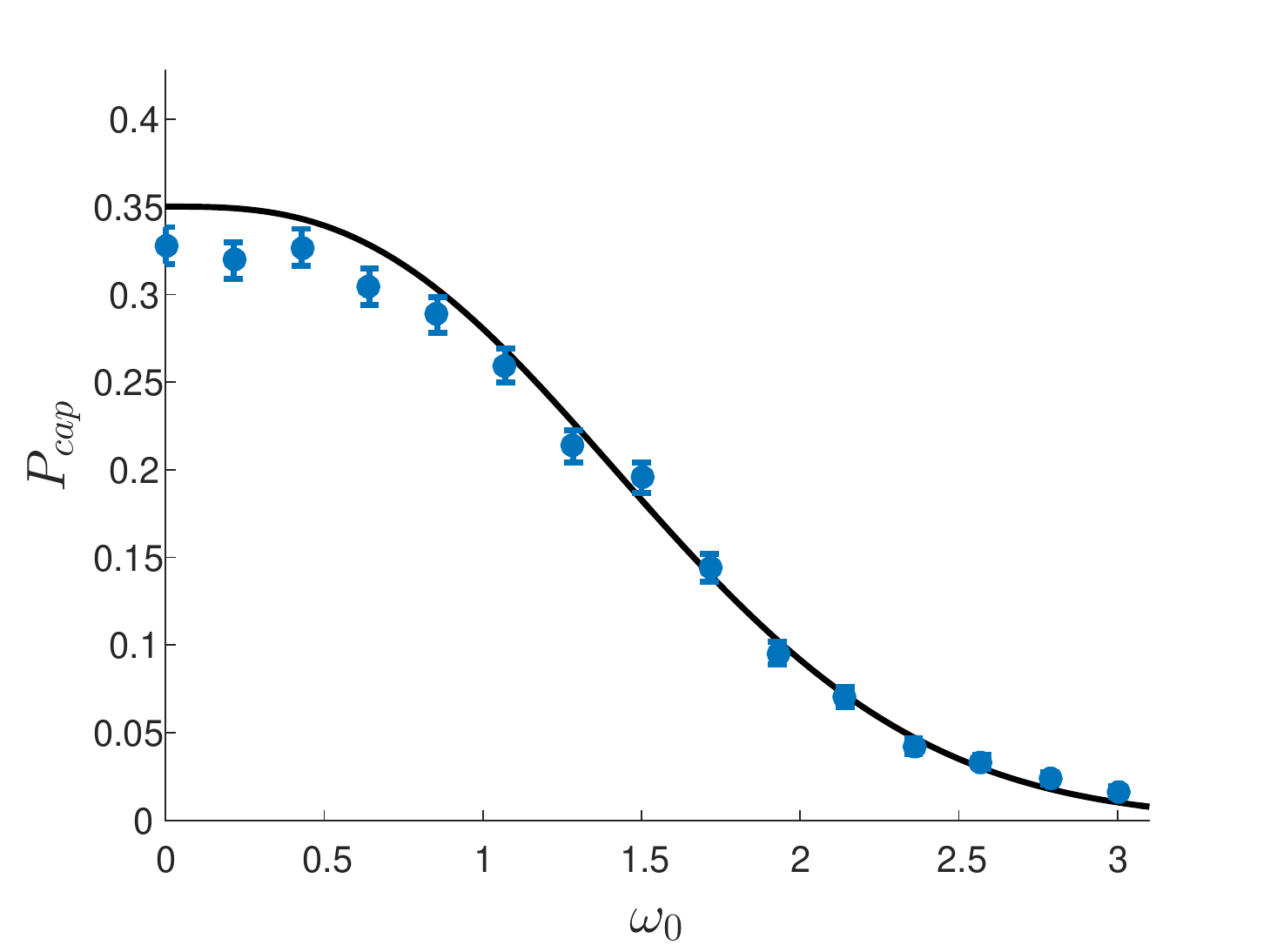}
\caption{Monte Carlo simulations (single resonance approximation) of the
resonant capture probability of initially thermal ensemble (2000 molecules),
versus the initial normalized driving frequency $\protect\omega _{0}$. The
solid line is the analytic result [see Eq. (\protect\ref{Pcap})]. The
parameters are $P_{1}=1.58$, $P_{2}=10$ and $\protect\omega_f=8$.}
\label{fig-h}
\end{figure}

We illustrate our theory in Fig. \ref{fig-h}, where the prediction of Eq. (%
\ref{Pcap}) is compared with numerical simulations (single resonance
approximation). We applied the OC drive to a thermal ensemble for parameters
$P_{1}=1.58$, $P_{2}=10$. The final normalized driving frequency in \ this
example was $8$, while the initial normalized driving frequency was varied.
One observes an excellent agreement of the theory (black line) with
simulations. Note that counter-intuitively, when $\omega _{0}$ decreases and
$P_{th}\left( \omega _{0}\right) $ becomes small, the capture probability
increases and reaches a maximum. In these cases, the vast majority of
captured molecules cross the separatrix during the evolution, and don't
start in resonance initially.

\begin{figure}[t]
\includegraphics[width=3.375in,clip=]{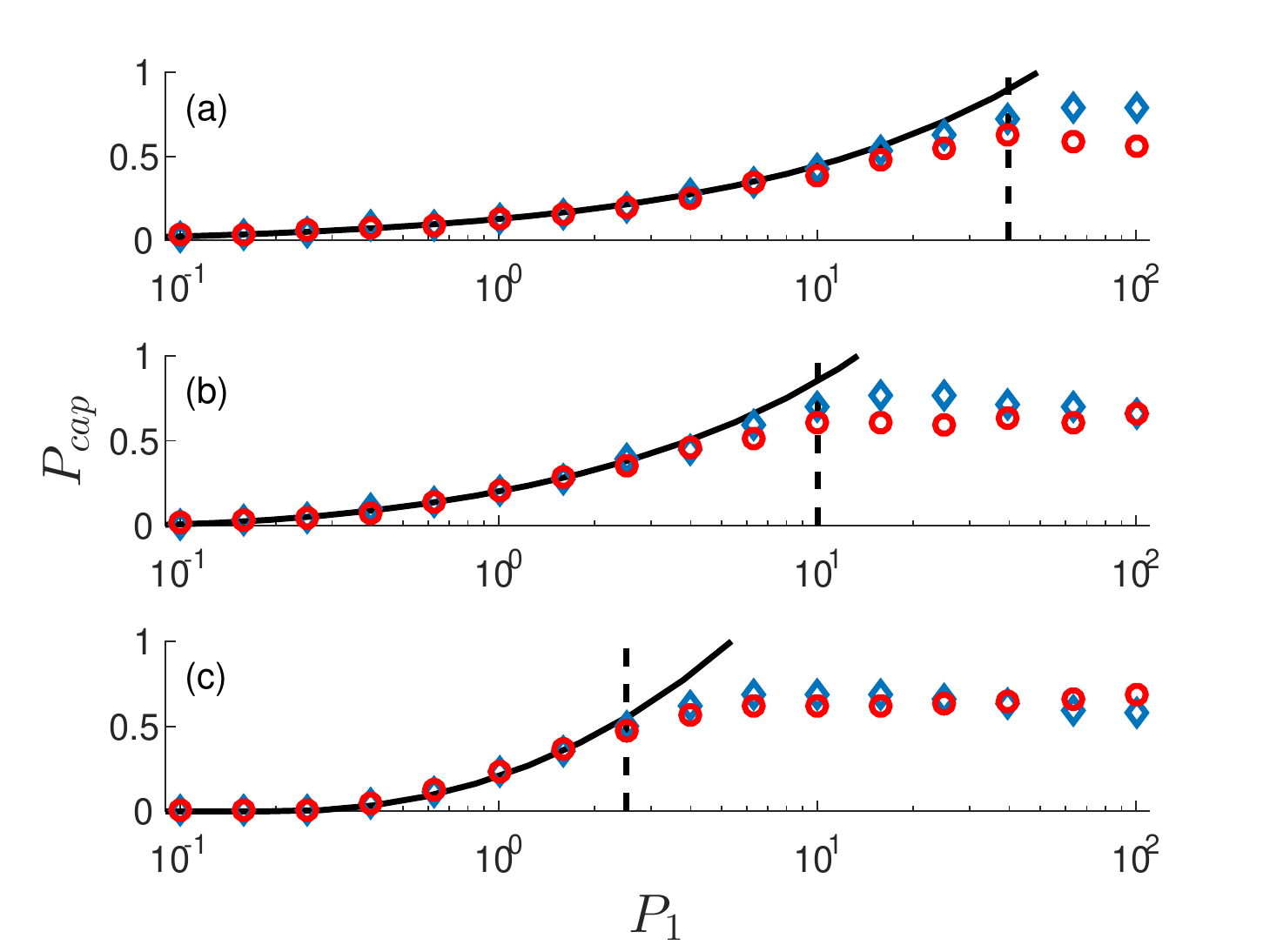}
\caption{(Color online) The resonant capture probability for three equal $%
P_2 $ lines in $P_{1,2}$ parameter space in Fig. \protect\ref{fig-e}. The
red circles show full simulation in original spherical coordinates, blue
diamonds are single resonance simulations, and analytic result is shown by
the solid line. The parameters are $P_2=39.8$ (a), $10$ (b) and $2.51$ (c), $%
\protect\omega_0=1$ and $\protect\omega_f=5$. Dashed lines show the location
of the weak drive limit. The number of molecules in simulations ranges from
500 to 5000, but the numerical uncertainty in all cases is smaller than the
marker size.}
\label{fig-f}
\end{figure}

Additional results are presented in Fig. \ref{fig-f}, testing a broader
range of parameters. In each panel in the figure, the OC drive with
normalized frequency varying from $\omega _{0}=1$ to $\omega _{f}=5$ is
applied to a thermal ensemble and $P_{2}$ is kept constant at $39.8$ (a), $%
10 $ (b) and $2.51$ (c), while $P_{1}$ is varied. The numerical results
include the simulations in spherical coordinates (blue diamonds), the single
resonance simulations (red circles), and both are compared with the
analytical result (solid line). One can see that the analytic prediction
correctly describes the simulations only in a certain range of parameters.
This is not surprising, as several approximations were made in the theory,
and need to be discussed next. One such approximation is the relative
smallness $\Delta L\ll 1$ of the width of the separatrix in $L$. In terms of
parameters $P_{1,2},$ this condition yields inequality%
\begin{equation}
\sqrt{P_{1}/P_{2}}\ll 1,  \label{small_delL}
\end{equation}%
which justifies the approximation in Eq. (\ref{SR4a}). In addition, we used
the single resonance assumption, allowing to discard higher nonresonant
harmonic contribution in deriving Eq. (\ref{HAASR}), which requires $%
P_{1}/P_{2}\ll 1$ and is guaranteed by (\ref{small_delL}). The location of $%
P_{1}=P_{2}$ is shown in Fig. \ref{fig-f} by dashed lines and one can see
that both types of simulations agree until one violates condition $%
P_{1}/P_{2}\ll 1$, but the theoretical curves deviate earlier, because
condition (\ref{small_delL}) is stricter. The ratio $P_{1}/P_{2}$ measures
the relative strength of the drive, so Eq. (\ref{small_delL}) describes the
weak drive limit.

\begin{figure}[t]
\includegraphics[width=3.375in,clip=]{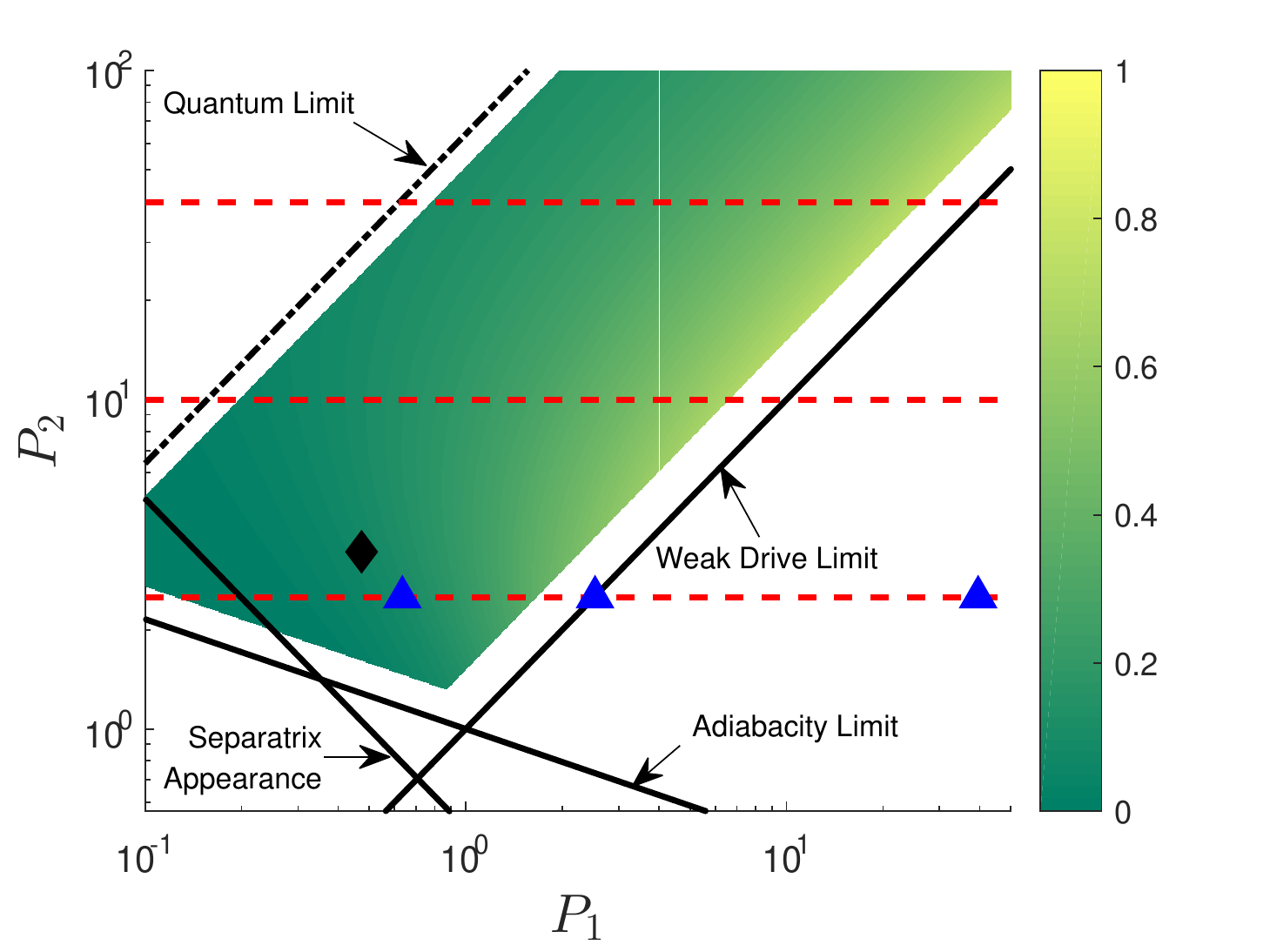}
\caption{(Color online) Validity conditions in $P_{1,2}$ parameter space.
The color coding represents the capture probability for a drive with $%
\protect\omega _{0}=1$, $\protect\omega_f=5$. The black lines are the weak
drive limit (Eq. \protect\ref{small_delL}), location of formation of
separatrix (Eq. \protect\ref{Separ}), and the adiabacity condition (Eq.
\protect\ref{adiab}). The dashed-dot black line is an example of the quantum
limit for $O_{2}$ at room temperature. The horizontal red dashed lines
represent the values of $P_{1,2} $ simulated in Fig \protect\ref{fig-f}.
The blue triangles are the parameters used in Fig. \protect\ref{fig-a},
while the red diamond shows parameters used in experiments \protect\cite%
{milner1,milner2,milner5}.}
\label{fig-e}
\end{figure}

Another assumption of the theory is the adiabaticity of autoresonant
evolution, i.e. $\nu ^{-2}d\nu /d\tau \ll 1$, where $\nu =\sqrt{4P_{1}P_{2}V}
$ is the characteristic frequency of autoresonant modulations (oscillations
of trajectories trapped inside the separatrix). We estimate $d\nu /d\tau
\sim O(\sqrt{P_{1}P_{2}}dL_{r}/dt)\sim O(\sqrt{P_{1}/P_{2}})$ and, therefore
the adiabaticity is guaranteed if
\begin{equation}
P_{2}P_{1}^{1/3}\gg 1.  \label{adiab}
\end{equation}%
Note that the resonant capture is impossible when there is no separatrix (no
trapped trajectories) for all $C$ values, which leads to the condition
\begin{equation}
P_{1}P_{2}>1/2  \label{Separ}
\end{equation}%
for trapping. While this condition doesn't affect the validity of the
results, it provides a useful border in $P_{1,2}$ parameter space. We
summarize this analysis in Fig \ref{fig-e} showing the $P_{1,2}$ parameter
space with boundaries defined by the above conditions as black solid lines
and the region of validity of the analytic results in color with the color
map corresponding to the theoretical capture probability for $\omega _{0}=1$
and $\omega _{f}=5$. The black diamond in the figure shows the conditions of
experiments \cite{milner1,milner2,milner5} ($E_{0}\approx 4.3\cdot 10^{9}\,%
\text{volt/m}$, $\beta \approx 1.7\times 10^{24}\,sec^{-2}$ for $O_{2}$
molecules at room temperature), which are in the region of validity of the
theory. The red dashed lines mark the parameter range in simulations in Fig. %
\ref{fig-f}, while the blue triangles show the conditions of simulations in
Fig. \ref{fig-a} with panel (c) in this figure way outside the weak drive
limit.

At this stage, we discuss the assumed classicality of our system. The
classical thermal distribution (\ref{TD-a}) is valid only when the most
probable $j$, the quantum number associated with the total angular momentum,
in the thermal equilibrium is large , say $j_{th}>5$. In addition, the
dynamics of trapped molecules must be classical. For this to be true, the
characteristic area $S$ (dimensional) of the separatrix in phase space must
exceed the Planck's constant $h$, so mixing of a few angular momentum states
would be possible. Then, the inequality $\sqrt{P_{2}/P_{1}}<j_{th}$ can
serve as a condition for classicality of trapped trajectories. An example of
this condition is presented in Fig. \ref{fig-e} by the dot-dashed line for $%
O_{2}$ at room temperature. Unlike the rest of the above conditions, this
line is not fixed in the $P_{1,2}$ space, and is both temperature and
molecule dependent via $j_{th}$ ($j_{th}=8$ in the figure). Note that the
classical results presented in this work are in the range of typical OC
experiments. Note also that the conservation law $L-L_{z}=const$\ in our
theory is the classical counterpart of the OC quantum selection rule $%
\left\vert j,m\right\rangle \rightarrow \left\vert j+2,m+2\right\rangle $,
where $m$ is the magnetic quantum number \cite{girard}.

Finally, in developing the theory, we have assumed that the characteristic
parameters $P_{1,2}$ are constant. In typical experiments these parameters
may vary in time. For example, the laser pulse amplitude may have slow
temporal dependence, the chirp rate $\beta $ may vary in time, and the
trapped molecules may experience slow centrifugal expansion at high rotation
speeds. Because of the adiabaticity, these effects can be taken into account
within our theory by using instantaneous values of $P_{1,2}$. For example,
the adiabaticity guarantees continued trapping in the system as long as $%
P_{1}P_{2}V$ (see Eq. \ref{SR7}) is an increasing function of time. \ If \
this function starts to decrease because of the aforementioned variation of
parameters, some molecules can escape the trapping. This effect of "leaked
molecules" was recently observed experimentally \cite{milner1,milner2}. Note
that this leakage can be stopped by slowly increasing the driving amplitude,
i.e. $P_{1}$ in time.


\section{SUMMARY}
\label{summary}
In conclusion, we have studied the process of capture of an
ensemble of molecules into resonance in the optical centrifuge and
calculated the associated capture probability. Based on three characteristic
time scales in the problem, we have introduced two dimensionless parameters $%
P_{1,2}$ (see Eqs. (\ref{P1}) and (\ref{P2})), transformed the problem to action-angle representation, and
applied the single resonance approximation in our analysis, allowing a significant acceleration of numerical simulations. We have then
studied the continuous phase space dynamics of the reduced one degree of
freedom system and found the probability of filling of separatrix by newly
trapped molecules. This calculation was based on the adiabaticity in the
problem and the incompressibility of the phase-space, avoiding the complex
issue of deciding the fate of individual trajectories. For a thermal
ensemble, we have compared the analytic results with numerical simulations,
showing excellent agreement, provided one satisfies the weak drive limit, the
adiabaticity and the classicality conditions, which were mapped in $P_{1,2}$ parameter space. It is shown that these conditions hold in current experimental setups. The results of this work can
be used in analysing existing and planning future experiments. It also seems important to generalize the theory into the quantum regime and study the transition from the quantum ladder climbing to the classical autoresonance \cite{AR6a,AR9} in the problem of molecular rotations. Finally, a similar phase space analysis can be applied in studying the problem of capture into autoresonance in other dynamical systems.

\begin{acknowledgments}
This work was supported by the Israel Science Foundation grant 30/14.
\end{acknowledgments}


\appendix

\section{}

\label{Appendix}

The transformation to action-angle variables discussed in Sec. (\ref{model})
is carried out similarly to \cite{Goldstein-trans,rydberg-atom}. We proceed by
solving the Hamilton-Jacobi equation in the problem, to obtain the
generating function \cite{Goldstein-trans}

\begin{equation}
W(L,L_{z},\varphi ,\theta )=\pm \int \sqrt{L^{2}-\frac{L_{z}^{2}}{\sin
^{2}\theta }}d\theta +\varphi L_{z},
\end{equation}%
where the actions are the angular momentum $L$ and its projection $L_{z}$ on
the $Z$-axix, the integration is along the trajectory, and the choice of the
sign accounts for the difference between the ascending and descending nodes.
The canonical transformation equations in this case are:

\begin{align}
p_{\varphi }& =\frac{\partial W}{\partial \varphi } & =& L_{z}, \\
p_{\theta }& =\frac{\partial W}{\partial \theta } & =& \pm \sqrt{L^{2}-\frac{%
L_{z}^{2}}{\sin ^{2}\theta }}, \\
\Theta _{L}& =\frac{\partial W}{\partial L} & =& \pm \int \frac{L\sin \theta
}{\sqrt{L^{2}\sin ^{2}\theta -L_{z}^{2}}}d\theta ,  \label{app1} \\
\Theta _{L_{z}}& =\frac{\partial W}{\partial L_{z}} & =& \mp \int \frac{L_{z}%
}{\sin ^{2}\theta }\frac{1}{\sqrt{L^{2}-\frac{L_{z}^{2}}{\sin ^{2}\theta }}}%
d\theta +\varphi .  \label{app2}
\end{align}%
It has been shown in \cite{Goldstein-trans} that the angles $\Theta _{L},\Theta
_{L_{z}}$ are two of the Euler angles, $\Theta _{L}$ measures the rotation
of the molecule in its plane of rotation, while $\Theta _{L_{z}}$ measures
the precession of the rotation plane itself. Substitution of the first two
transformation equations into the unperturbed Hamiltonian yields
\begin{equation}
H_{0}=P_{2}L^{2}/2.
\end{equation}%
For calculating the perturbed part of the Hamiltonian we set $\Theta _{L}=0$
when $\theta $ is at its minimal value, and $\Theta _{L_{z}}=0$ when the
line of nodes is along the $X$ axis, and solve the integrals in Eq. (\ref%
{app1}), (\ref{app2}) to find:

\begin{align}
\cos {\theta }& =\sqrt{1-L_{z}^{2}/L^{2}}\cos {\Theta _{L}},  \label{app4} \\
\varphi & =\Theta _{L_{z}}+\arctan [{(L/L_{z})\tan {\Theta _{L}]}}+\frac{\pi
}{2}.  \label{app4a}
\end{align}%
\textbf{\ }Next, we define $s=signL_{z}$ and notice that $\arctan {\left(
L/L_{z}\tan {\Theta _{L}}\right) }$ can be written as the sum $s\Theta
_{L}+sf\left( \left\vert L/L_{z}\right\vert ,\Theta _{L}\right) $, where $f$
is a periodic function of $\Theta _{L}$ of period $\pi $. We expand this
function in Fourier series to get

\begin{equation}
f\left( \left\vert L/L_{z}\right\vert ,\Theta _{L}\right)
=\sum_{n=1}^{\infty }\frac{\left( \left\vert L/L_{z}\right\vert -1\right)
^{n}}{n\left( \left\vert L/L_{z}\right\vert +1\right) ^{n}}\sin {\left(
2n\Theta _{L}\right) },
\end{equation}%
which, in terms of $A=\left( \left\vert L/L_{z}\right\vert -1\right) /\left(
\left\vert L/L_{z}\right\vert +1\right) ,$ becomes:

\begin{equation}
f\left( \left\vert L/L_{z}\right\vert ,\Theta _{L}\right) =-\frac{i}{2}\log
\left( \frac{1-Ae^{-2i\Theta _{L}}}{1-Ae^{2i\Theta _{L}}}\right) .
\label{app5}
\end{equation}%
At this point, we write the action-angle representation of the perturbed
part of the Hamiltonian using Eqs. (\ref{app4}), (\ref{app4a}):

\begin{equation}
U=\left[ \left( 1-L_{z}^{2}/L^{2}\right) \cos ^{2}\Theta _{L}-1\right] \sin
^{2}\Psi ,  \label{ww}
\end{equation}%
where $\Psi =\Theta _{L_{z}}+s\Theta _{L}+sf-\phi _{d}$ and then use Eq. (%
\ref{app5}) to find the closed form expressions for $\cos \left( 2f\right)
,\sin \left( 2f\right) $. We define the phase mismatch $\Phi =2(\Theta
_{L}+\Theta _{L_{z}}-\phi _{d})$ in the problem, use this definition to
replace $2(\Theta _{L_{z}}-\phi _{d})$ in (\ref{ww}) and average the
resulting $U(L_{z}/L,\Theta _{L},\Phi )$ with respect to the fast phase $%
\Theta _{L}$. This yields the full Hamiltonian in the single resonance
approximation:
\begin{equation}
H\left( \Theta _{L},\Theta _{L_{z}},L,L_{z}\right) \approx P_{2}\frac{L^{2}}{%
2}+P_{1}V\cos {\Phi }+P_{1}F,
\end{equation}%
where
\begin{eqnarray}
V &=&\frac{1}{8}\left( 1+\frac{L_{z}}{L}\right) ^{2}, \\
F &=&\frac{1}{4}\left( 1-\frac{{L_{z}}^{2}}{L^{2}}\right) .
\end{eqnarray}%
Note that this result is independent of $s$ and that angle $\Theta _{L}$
exhibits non-trivial behavior, as it always increases, regardless the
direction of rotation (given by $s$).


\end{document}